\documentclass[reprint,amsmath,amssymb,aps,prl]{revtex4-1}

\usepackage{graphicx}
\usepackage{dcolumn}
\usepackage{bm}
\usepackage[usenames]{color}

\begin{document}

\preprint{BAP'S/123-QED}

\title{Finite-temperature transition of the distorted kagome-lattice Heisenberg antiferromagnet}

\author{Hiroshi Masuda }
\author{Tsuyoshi Okubo}
\author{Hikaru Kawamura}
 \email{kawamura@ess.sci.osaka-u.ac.jp}
\affiliation{Department of Earth and Space Science,
 Graduate School
 of Science, Osaka University, Toyonaka, Osaka 560-0043, Japan%
}%
\date{\today}

\begin{abstract}
Motivated by the recent experiment on kagome-lattice antiferromagnets, we study the zero-field ordering behavior of the antiferromagnetic classical Heisenberg model on a uniaxially distorted kagome lattice by Monte Carlo simulations. A first-order transition, which has no counterpart in the corresponding undistorted model, takes place at a very low temperature. Origin of the transition is ascribed to a cooperative proliferation of topological excitations inherent to the model.
\end{abstract}

\maketitle

 Renewed interest has recently been paid to novel ordering properties of geometrically frustrated magnets \cite{Journal_topic,Book}. Two-dimensional (2D) kagome lattice, a corner-sharing network of regular triangles, is a typical example of such geometrically frustrated lattices. Many theoretical works have been performed to elucidate the ordering  of the antiferromagnetic (AF) Heisenberg model on this lattice for both cases of quantum $S=1/2$ \cite{Waldtman,Singh,Jiang,NakanoSakai} and classical $S=\infty$  \cite{Chalker,Reimers,Zhitomirsky1,Zhitomirsky2,Robert} spins.

 In the quantum $S=1/2$ case, although there seems to be a consensus among researchers that the ground state of the model lacks in the conventional AF long-range order, there still exists a considerable diversity in the view about the nature of its ground state \cite{Waldtman,Singh,Jiang,NakanoSakai}.
 In the classical $S=\infty$ case, any combination of the local 120-degrees spin structure on a constituent triangle is allowed as its ground state, leading to the massive degeneracy. The effect of ``order from disorder'' is operative at low temperatures, which favors at the harmonic level the coplanar states, or the spin nematic states, where all spins are contained on a common plane in the spin space \cite{Chalker}. Yet, the coplanar ground states themselves are heavily degenerate, the degeneracy of $O(e^N)$ ($N$ the number of spins). In order to determine which coplanar state is realized in the $T\rightarrow 0$ limit, nonlinear excitations beyond the harmonic level need to be invoked. The model selects among them the so-called $\sqrt 3\times \sqrt 3$ state \cite{Reimers}.

 Experimentally, there have been few realizations of Heisenberg kagome AFs. Only recently, several possible realizations of $S=1/2$ Heisenberg kagome AFs were reported, {\it e.g.\/}, herbertsmithite \cite{Nocera,Mendelse}, volborthite \cite{HYoshida,MYoshida,Nakazawa} and vesigniete \cite{Okamoto}. Volborthite Cu$_3$V$_2$O$_7$(OH)$_2\cdot$2H$_2$O and vesigniete BaCu$_3$V$_2$O$_8$(OH)$_2$
are structurally distorted from the perfect kagome lattice in a uniaxial manner where the regular triangle is distorted to isosceles triangles. This structural distortion gives rise to two distinct exchange couplings, $J_1$ and $J_2$, as demonstrated in the inset of Fig.1(b), where the extent of the distortion may be represented by the parameter $r=J_2/J_1$. Whether $r$ is greater or smaller than unity in volborthite has not fully been determined experimentally, though the bond distance seems to favor $r>1$ \cite{YBKim}. An interesting recent experimental finding is that this compound exhibits a thermodynamic phase transition at $T_c\simeq 1$K, with a slowly fluctuating ordered state \cite{HYoshida,MYoshida,Nakazawa}. 

 Under such circumstances, it remains most interesting to clarify what type of ordering behavior is expected in the kagome Heisenberg AF {\it under uniaxial distortion\/}. In the present paper, we address this issue for the classical model in zero field. In particular, we wish to clarify what type of spin structure is chosen in the $T\rightarrow 0$ limit, and whether a finite-temperature transition is ever possible.

 Only a few theoretical works have been made on the AF classical Heisenberg model on the distorted kagome lattice in the past. Y.B. Kim {\it et al\/} studied the model by an effective ``chirality'' Hamiltonian approach, and suggested that the so-called ``chirality-stripe'' state might be selected in the low-temperature limit \cite{YBKim}. Kaneko {\it et al\/} studied the in-field properties of the model by Monte Carlo (MC) simulations, and observed a field-induced phase transition with a weak (zero)-field state being the $\sqrt 3 \times \sqrt 3$ state \cite{Kaneko}.

 In the present paper, we study the zero-field ordering properties of the model by means of extensive MC simulations, paying particular attention to the low-temperature regime which was not examined in previous works. Surprisingly, we find a clear first-order transition, which has no counterpart in the undistorted model, in the extremely low-temperature regime.

 The model we consider is the classical AF  Heisenberg model on a uniaxially distorted kagome lattice with two distinct AF nearest-neighbor couplings $J_1$ and $J_2=r J_1$ as shown in the inset of Fig.1(b). The Hamiltonian is given by

\begin{equation}
 \mathcal{H}= J_1\sum_{\left\langle i,j
\right\rangle}\bm{S}_i\cdot\bm{S}_j +J_{2}\sum_{\langle\langle i,j
\rangle\rangle}\bm{S}_i\cdot\bm{S}_j.
\label{Hamiltonian}
\end{equation}

 In the undistorted case $r=1$, the ground-state of a single triangle is the 120-degrees spin structure, whereas, in the distorted case $r\neq 1$, it is distorted such that a spin-canting angle at the $J_1$-bond becomes $\theta =\arccos(-\frac{1}{2r})$. The ground state of the distorted model is highly degenerate in that any combination of the local ground state on a triangle yields a ground state of the entire lattice \cite{YBKim,Kaneko}. Order from disorder effect due to harmonic excitations then selects coplanar states as in the undistorted case. The coplanar states are still heavily degenerate, but the amount of degeneracy is much reduced from that in the undistorted case, from the extensive number of $O(e^N)$ to the subextensive number of $O(e^L)$ ($L$ the linear dimension) \cite{YBKim}. It should be noticed that the $\sqrt 3\times \sqrt 3$ state, which is the $T\rightarrow 0$ state of the undistorted model, cannot be a ground state of the distorted model any more. 

 In order to determine which state is favored in the $T\rightarrow 0$ limit in the distorted model, we perform an extensive MC simulation at low temperatures. Our MC simulation is a combination of the heat-bath and the over-relaxation methods. A unit MC sweep consists of single heat-bath sweep and subsequent 10 over-relaxation sweeps. Various lattice sizes are studied up to $L=192$ with total $N=3\times L^2$ spins. We generate $1\times 10^6\sim 4\times 10^7$ MC steps per spin (MCS) at each temperature, both cooling and warming runs being made. For our check of equilibration, see the online Supplemental Material \cite{equilibration}. We apply several types of boundary conditions (BC), including periodic, free, ``fixed'' and ``vortex'' boundaries (to be explained below). Most extensive calculations are performed for $r=1.1$, while other values of $r>1$ are also studied.

\begin{figure}
  \includegraphics[width=8cm]{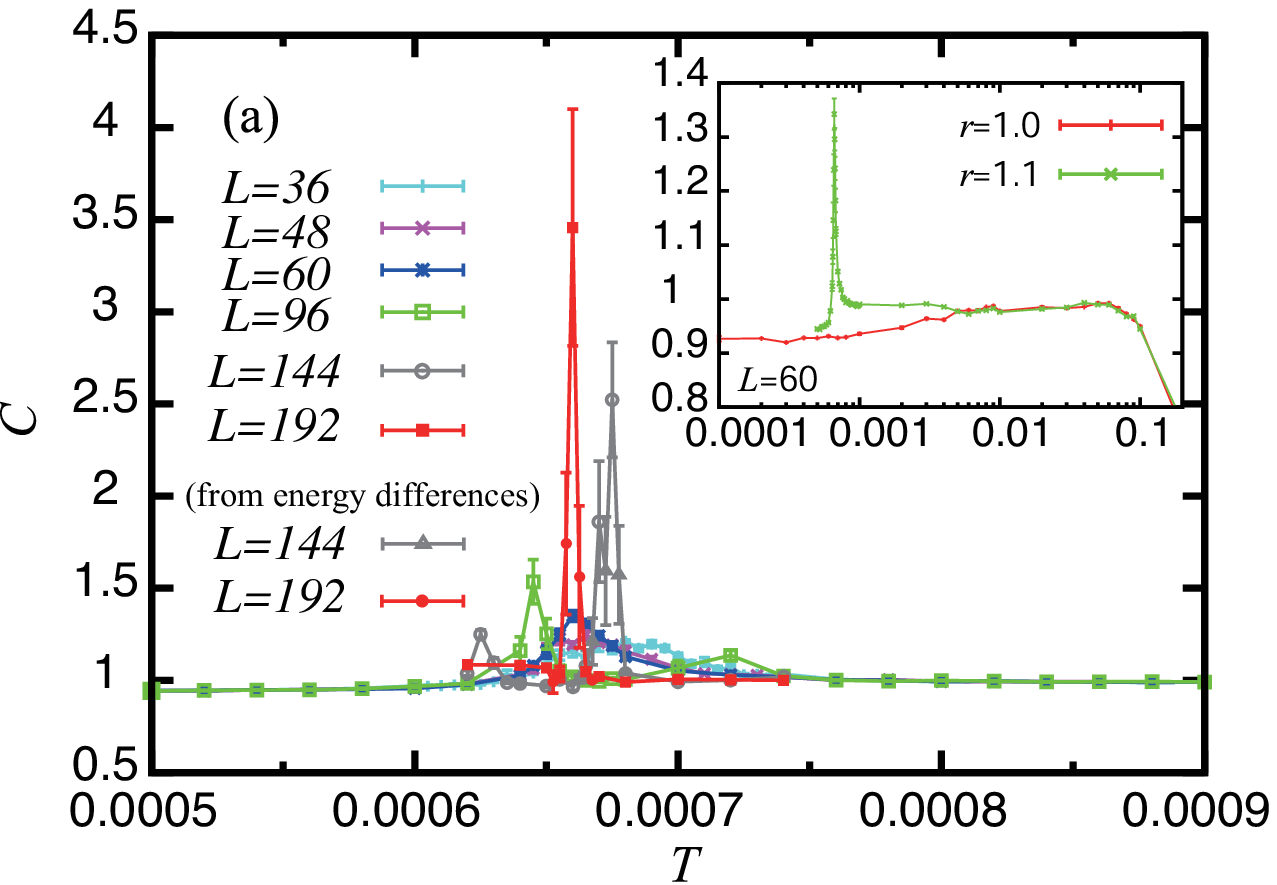}
  \includegraphics[width=8cm]{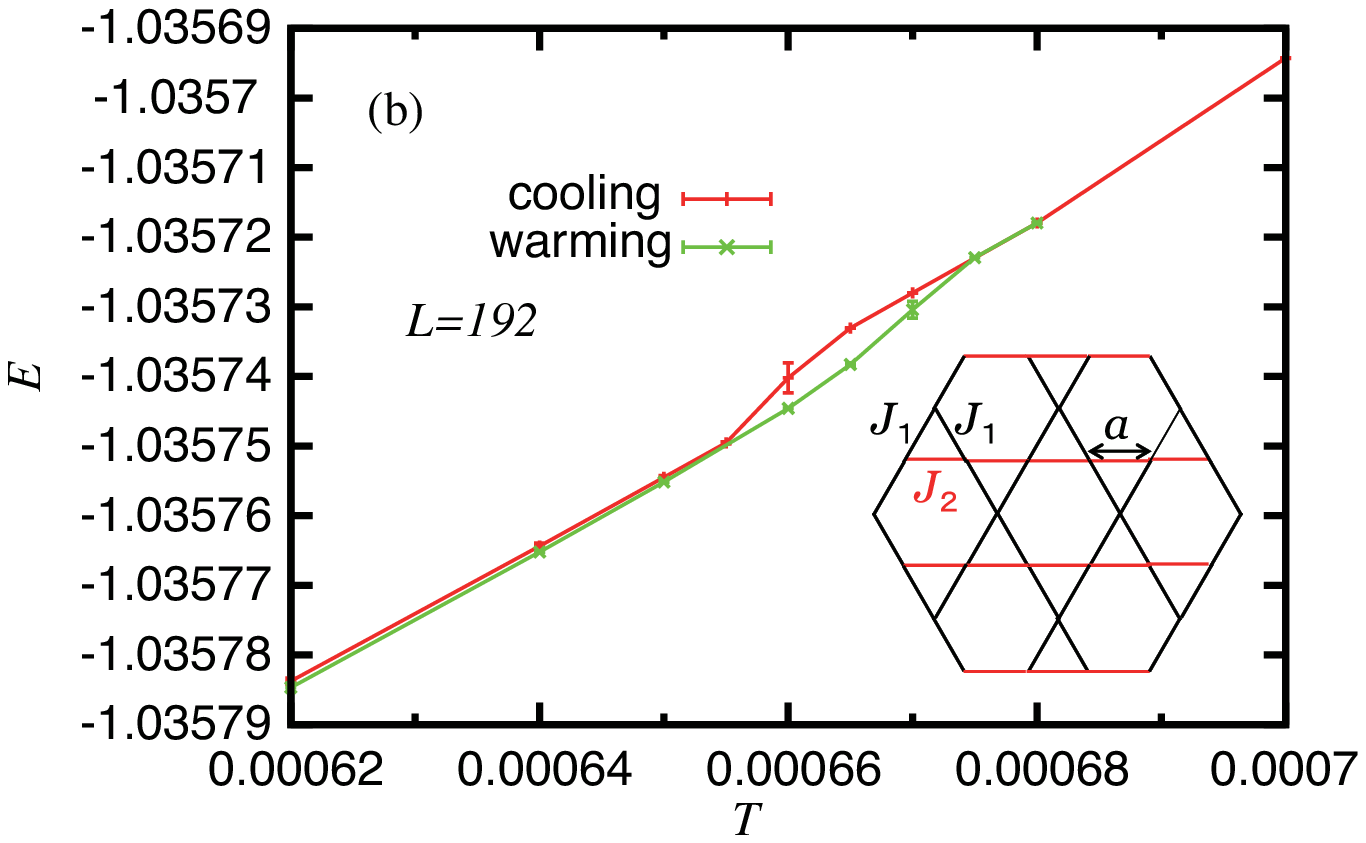}
 \caption{(color online). (a) The temperature and size dependence of the specific heat per spin for the cooling run of the distorted model of $r=1.1$ under periodic BC, either calculated from the energy fluctuation or from the temperature difference of the energy. In the inset, the data for $L=60$ are compared with the corresponding data of the undistorted model $r=1$ in a wider temperature range. (b) The temperature dependence of the energy of the distorted model of $r=1.1$ for $L=192$ both for the cooling and warming runs. The inset illustrates a distorted kagome lattice. 
}
 \label{fig_energy}
\end{figure}

 In Fig.1(a), we show the temperature dependence of the specific heat for the cooling runs for the case of $r=1.1$. In the inset, the data are compared with those of the undistorted model $r=1$. A sharp peak absent in the undistorted case appears, growing rapidly with  $L$. The data turn out to be reversible for smaller sizes of $L\leq 96$, while a weak hysteresis is observed for larger sizes of $L\geq 144$. In Fig.1(b), we show the temperature dependence of the energy for $L=192$, where a clear hysteresis is observed in the transition region signaling the first-order nature of the transition. This is consistent with the sharp peak of the specific heat growing rapidly with $L$. The first-order transition is extremely weak in that the latent heat is only $\Delta E \simeq 6\times 10^{-6}$ (in units of $J_1$). It is remarkable that such a small energy scale is generated in the model which has only $O(1)$ energy scale. 
A closer look of the data reveales that the specific heat exhibits, in addition to the sharp main peak, additional small peaks (or weak structures) at $L$-dependent temperatures. These small peaks are associated with a discontinuous shift of the $q$-value characterizing an incommensurate spiral in finite systems under periodic BC, and are expected to vanish in the thermodynamic limit: See below.

\begin{figure}
  \includegraphics[width=8cm]{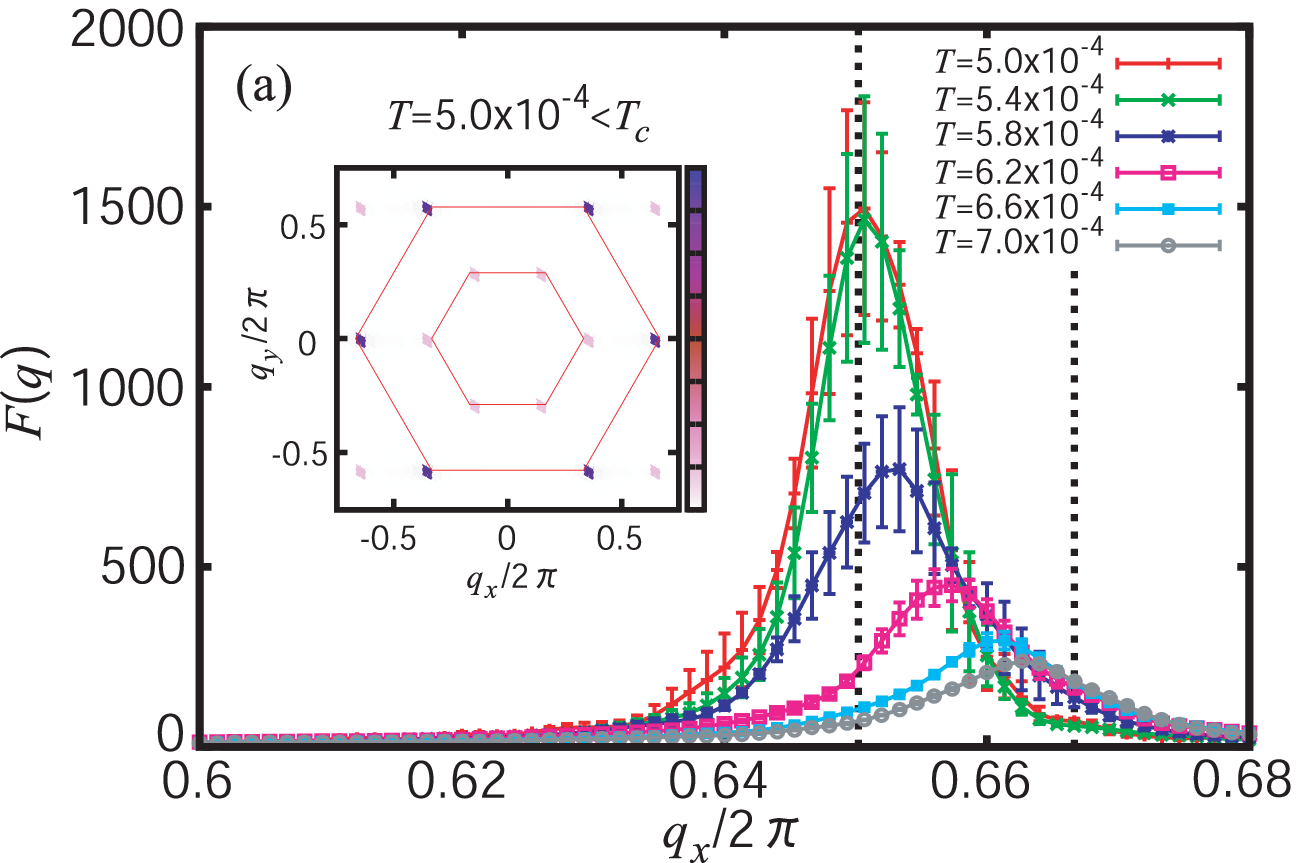}
  \includegraphics[width=8cm]{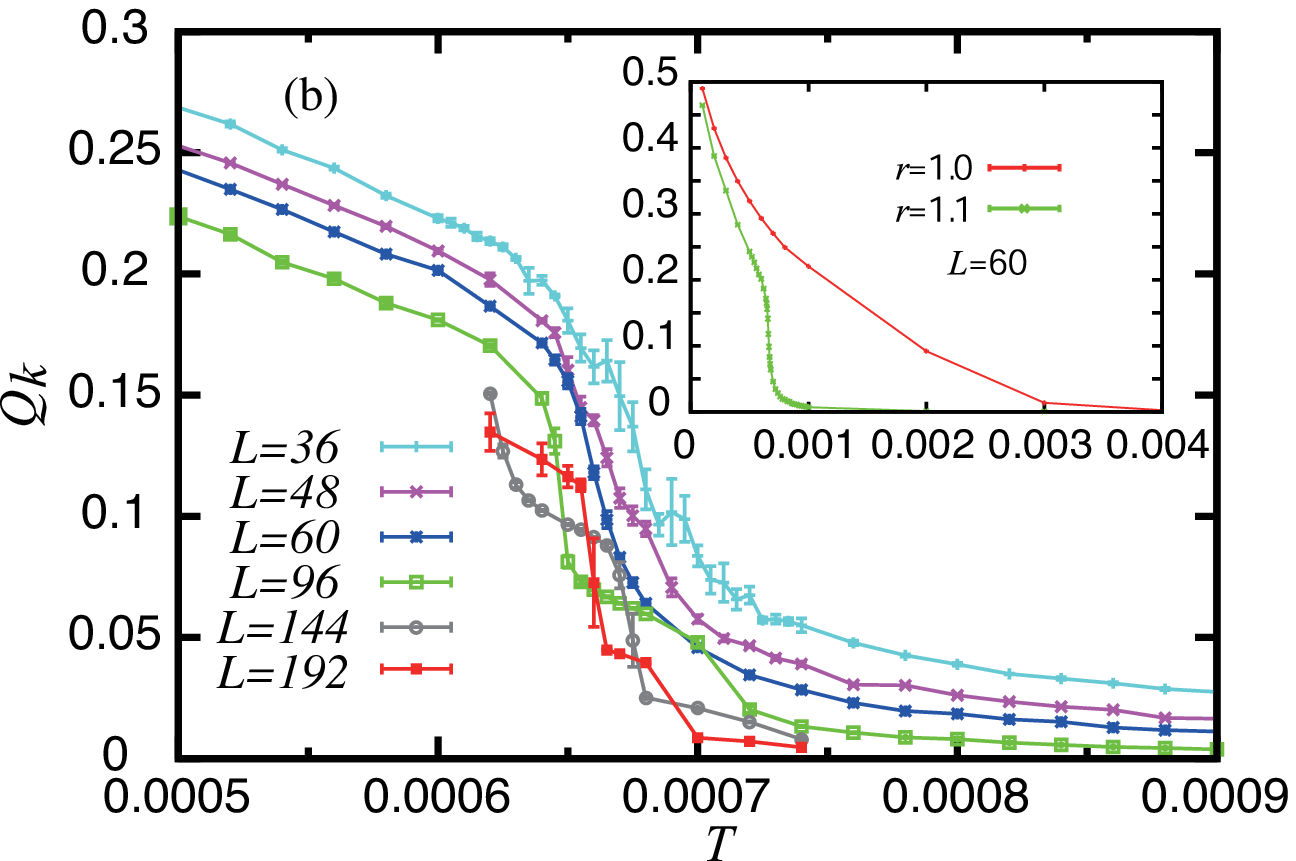}
 \caption{(color online). 
(a) The inset exhibits the spin structure factor in the wavevector ($q_x, q_y$) plane of the distorted model of $r=1.1$ at a temperature below $T_c$, $T=0.0005$. Outer (inner) hexagon represents the first Brillouin zone of the triangular lattice of a lattice constant $a$ ($2a$). Main panel plots $F(q_x, 0)$ versus $q_x$ in the vicinity of the $K$ point $q_x/(2\pi)=2/3$ for several temperatures across $T_c$. The lattice size is $L=96$ with free BC. The vertical dotted lines represent the $K$-point wavevector (right) and the incommensurate wavevector at $T=0$ (left). (b) The temperature and size dependence of the nematic order parameter of the distorted model of $r=1.1$ under periodic BC. In the inset, the $L=60$ data are compared with the corresponding data of the undistorted model of $r=1$.
}
 \label{fig_structure}
\end{figure}

 The inset of Fig.2(a) exhibits the spin structure factor $F$ in the $(q_x,q_y)$ plane calculated at a temperature below the bulk trnasition temperature $T_c\simeq 0.00065$, where $q$ is measured in units of $2\pi /a$ ($a$ the nearest-neighbor distance of the kagome lattice). The strongest intensity appears at a wavevector along the $J_2$-bond direction slightly off the $K$ point, corresponding to an incommensurate spiral state. The observed peak has only a finite width reflecting a finite spin correlation length of the 2D Heisenberg model at finite $T$. Such an incommensurate spiral state  is certainly a ground state of the distorted model, which reduces to the $\sqrt 3\times \sqrt 3$ state  in the undistorted limit $r\rightarrow 1$. In the main panel of Fig.2(a), we plot $F(q_x, 0)$  versus $q_x$ in the vicinity of the main peak for several temperatures across $T_c$. Free BC are applied here in order to minimize the finite-size effect associated with the incommensurability.  As one approaches $T_c$, the peak position gradually shifts from the $K$ point to an incommensurate position. This shift of the peak position is observed already above $T_c$ for larger $L$. When one applies periodic BC, the ensuing discretization of $q$ in units of $2\pi /L$ often hampers the observation of such a shift for small $L$. This might explain the reason why the peak was located just at the $K$ point in the simulation of Ref.\cite{Kaneko}. For relatively large but finite $L$, such a discretization effect of the periodic BC causes, with varying the temperature, a sudden jump in the $q$-value. In any case, the low-temperature state of the model turns out to be  an incommensurate spiral state, distinct from the chiral-spiral state inferred in Ref.\cite{YBKim}, or from the $\sqrt 3\times \sqrt 3$ state suggested in Ref.\cite{Kaneko}.

\begin{figure}
  \includegraphics[width=8cm]{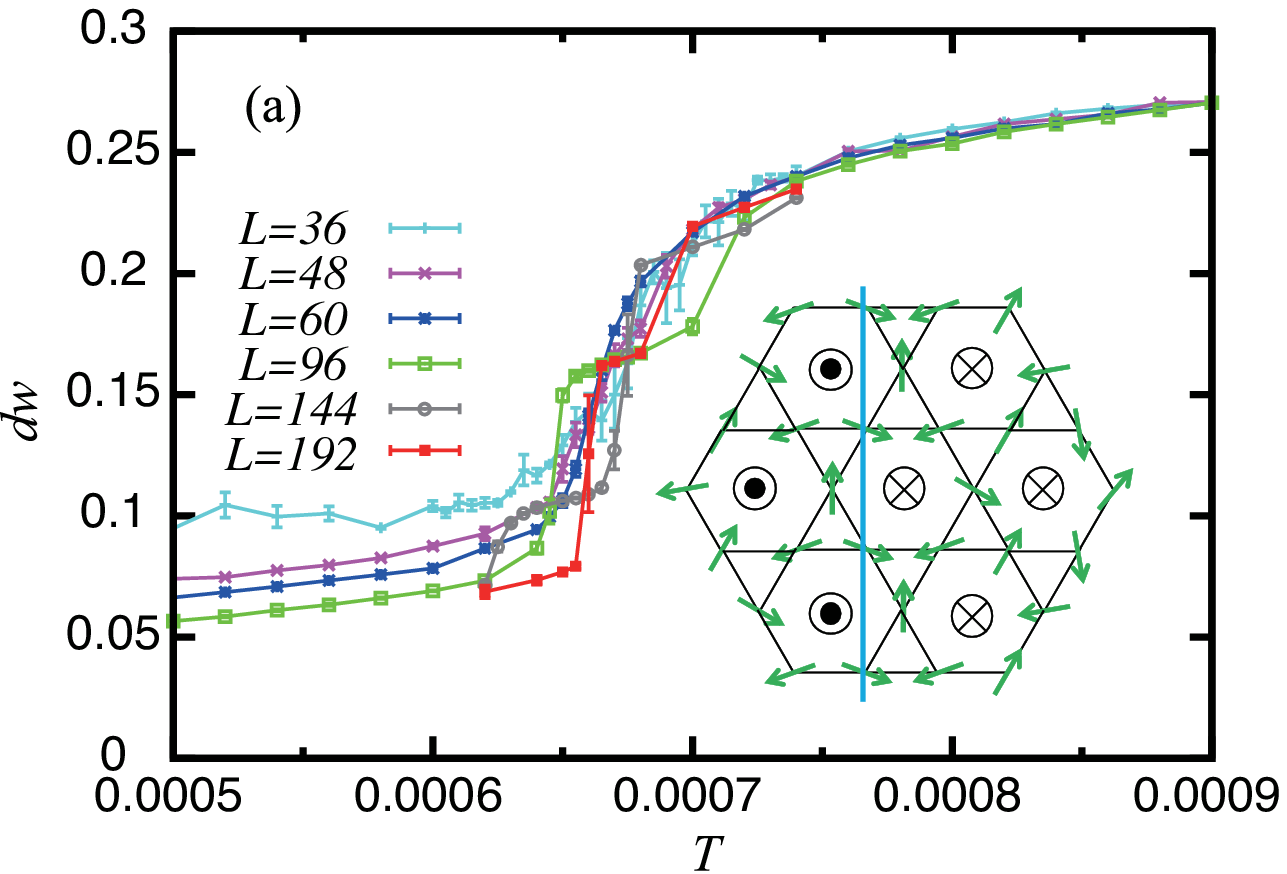}
  \includegraphics[width=8cm]{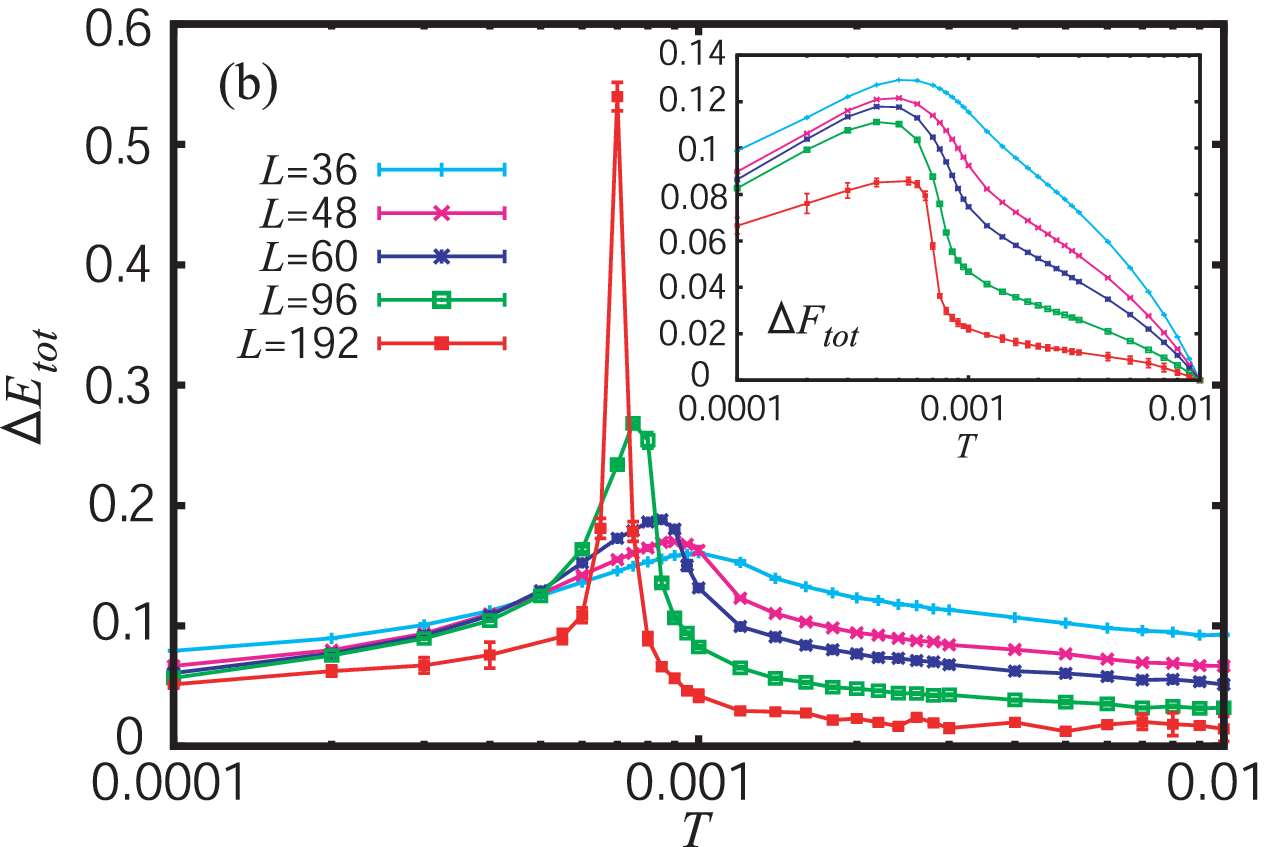}
 \caption{(color online). 
(a) The temperature and size dependence of the number density of chiral domain walls of the distorted model of $r=1.1$ under periodic BC. Inset exhibits a typical  chiral-wall configuration. The symbols represent the upward and downward chirality vectors. (b) The temperature and the size dependence of the vortex energy $\Delta E_{tot}(T)$   (the vortex free energy $\Delta F_{tot}(T)$) in the main panel (in the inset).
}
 \label{fig_nematic}
\end{figure}

In Fig.2(b), we show the temperature dependence of the nematic order parameter $Q_\kappa$ representing the extent of the spin coplanarity
\begin{equation}
Q_\kappa = \sum _{\mu \nu} \left( \frac{1}{N_t}\sum _\alpha Q_{\alpha \mu\nu} \right)^2, \ \ 
Q_{\alpha \mu\nu} = \kappa_{\alpha \mu}\kappa_{\alpha \nu}-\frac{1}{3}\vec \kappa_{\alpha}^2 \delta_{\mu\nu},
\end{equation}
where $N_t$ is the number of upward triangles and the vector chirality $\vec \kappa_\alpha$  at each upward triangle $\alpha$ is given by 
%
$
\vec \kappa = \frac{2}{3\sqrt{3}} (\vec S_1\times \vec S_2 + \vec S_2\times \vec S_3 + \vec S_3\times \vec S_1), 
$
where three spins $\vec S_1\sim \vec S_3$ are arranged in a counter-clockwise manner on a triangle.

 In any 2D Heisenberg model, the nematic order parameter should vanish at any finite $T$ in the thermodynamic limit. Yet, it gives a measure of the nematic short-range order (SRO). As can be seen from Fig.2(b), the nematic order or the spin coplanarity sets in almost simultaneously with the first-order transition for the case of $r=1.1$. Additional step-wise structures of the data are due to the jump in the $q$-value associated with the incommensurability effect in finite systems under periodic BC mentioned above. This onset of the nematic SRO occurs at a temperature significantly lower than that of the undistorted model: See the inset.

 Establishing the existence of a first-order transition, we further examine what is responsible for this transition. For this purpose, we study the behavior of two types of topological excitations inherent to the model. One is a {\it chiral-domain wall\/}, and the other is a {\it $Z_2$ vortex\/}. 

 The chiral domain wall is a wall-like excitation, which might be obtained by connecting  the reversal points of the vector chirality. It is a zero-mode arising from the coplanar nature of the low-$T$ spin state. In the present distorted model, the chiral wall becomes a zero-mode only for a straight-line wall of system size $L$ running perpendicular to the $J_2$-bond direction, in sharp contrast to the undistorted case where chiral walls of various sizes and shapes are possible including closed loops. 

  Fig.3(a) exhibits the temperature dependence of the total number density of chiral walls. Here we assign the vector chirality at each nearest-neighbor bond ($ij$) along the $J_2$-bond direction as $\vec \kappa_{ij}=\vec S_i\times \vec S_j$, and suppose that a lattice site $i$ is included in the chiral wall when the condition $\vec \kappa_{i,i+\hat e_x} \cdot \vec \kappa_{i-\hat e_x,i} < 0$ is met. As the temperature is lowered across $T_c$, the number of chiral walls diminishes rapidly. Typical snapshots of the chiral wall configurations are given in the online Supplemental Material \cite{wall}.

 We also study the behavior of another topological excitation, $Z_2$ vortex \cite{vortex}, a point defect inherent to the frustrated 2D Heisenberg model with the noncollinear spin order \cite{KawaMiya,Kawamura1,Kawamura2}. For this purpose, we introduce two different BC: One is the fixed BC where we fix the boundary spins so that they match the expected ground-state spin configuration of an incommensurate spiral lying on the ($S_x,S_y$) plane. The other is the ``vortex'' BC where the boundary spins are fixed in the manner to accommodate a single vortex on top of the fixed-BC spin configuration, {\it i.e.\/}, a spin rotation of $2\pi$ around the $S_z$-axis is applied to the boundary spins \cite{KawamuraKikuchi}. Then, we calculate the energy difference $\Delta E_{tot}(T)$ between the total energies of the fixed and the vortex BC, together with the total free-energy difference $\Delta F_{tot}(T)$ evaluated by integrating $\Delta E_{tot}(T)$ with respect to $T$ \cite{KawamuraKikuchi}. Fig.3(b) exhibits $\Delta E_{tot}(T)$ calculated in this way. A sharp peak is observed just at $T_c$, indicating that a strong anomaly occurs  at $T_c$ in the vortex sector. The calculated free-energy difference $\Delta F_{tot}(T)$ is shown in the inset, which indicates that the vortex tends to stiffen below $T_c$. 

 Thus, the transition turns out to accompany a significant change in the state of topological excitations, {\it i.e.\/}, accompany a proliferation (or a rapid increase in  number) of both chiral walls and $Z_2$ vortices. Indeed, as demonstrated in the online Supplemental Material \cite{cooperative}, there is an effective {\it attractive\/} interaction acting between the chiral domain walls and the $Z_2$ vortices, which might cause a cooperative thermal generation of these topological excitations possibly leading to the observed first-order transition. The $Z_2$-vortex transition often becomes of first order when it accompanies a simultaneous symmetry-breaking of other degrees of freedom \cite{Domange,Tamura}.

\begin{figure}
  \includegraphics[width=8cm]{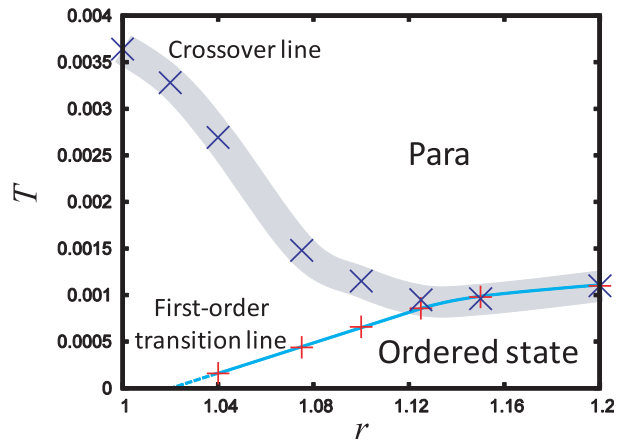}
 \caption{(color online). 
The phase diagram of the distorted kagome Heisenberg model in the temperature - distortion ($r$) plane. The undistorted model corresponds to $r=1$. The crossover line is conventionally drawn at a point where $Q_\kappa$ for $L=48$ is equal to 0.01. The asymptotic behavior of the transition line near the undistorted limit is presently not clearly determined. 
}
 \label{fig_phasediagram}
\end{figure}

 We have performed similar simulations for other values of $r>1$, and the resulting phase diagram is given in Fig.4 in the distortion $r$ vs. the temperature $T$ plane. Together with the first-order transition line $T_c(r)$, the crossover line associated with the onset of the nematic order is also shown. Interestingly, this crossover line coincides with the first-order transition line for $r\gtrsim 1.1$, while it is separated from the $T_c$-line for $r\lesssim 1.1$, lying well above it. For smaller $r<1.04$, we cannot identify the transition directly. This is either the transition itself is absent for smaller $r$, or $T_c$ becomes so low that we simply cannot observe it in our simulations. In the range $1.04 \lesssim r \lesssim 1.12$, $T_c$ varies almost linearly with $r$. Linear extrapolation yields a critical value $r_c\simeq 1.02$ below which we no longer have a finite-$T$ transition. However, the difficulty to get reliable data at low temperatures in this small $r$ region hampers us to reliable determine the fate of the transition line in the $r\rightarrow 1$ limit, {\it i.e.\/}, whether i) it hits the $T=0$ axis at $r=r_c>1$, or ii) shows a nonlinear behavior in the smaller $r$ region reaching the $T=0$ axis only at $r=1$, or iii) ends at a finite $T$ exhibiting a critical end point at ($r_c>1$, $T_c>0$). 

 Finally, we discuss the possible relation of the finite-$T$ transition observed in our simulation to the zero-field transition of volborthite. The  Curie temperature of the present model is estimated to be $T_{CW}\simeq 0.9$ for $r=1.1$. If one matches it with the corresponding experimental value of volborthite $\sim 115$K \cite{Hiroi}, $T_c$ is estimated to be $0.1$K, which is an order of magnitude smaller than the experimental value, 1K. Several possible causes for this discrepancy are conceivable: i) The experimental value of the distortion $r$ is largely unknown which could explain at least a part of the discrepancy. ii) Volborthite is a $S=1/2$ system. Quantum effect neglected here might give significant correction. iii) Real material is 3D. Weak interplane coupling neglected here might push up the $T_c$-value significantly. iv) The Dzyaloshinskii-Moriya interaction neglected here might change the ordering behavior significantly. Further studies are then required to clarify the true relation between the transition found here and the experimental one. 

 In summary, we studied the ordering of the classical Heisenberg AF on a distorted kagome lattice by MC simulations. We have found that the model exhibits a thermodynamic first-order transition at an extremely low but finite temperature. Cooperative generation of the chiral domain walls and the $Z_2$ vortices might be responsible for this first-order transition.

\acknowledgments

The authors are thankful to Z. Hiroi, H. Yoshida, and M. Yoshida for
 useful discussion. This study was supported by Grand-in-Aid for
Scientific Research on Priority Areas ``Novel States of Matter Induced
by Frustration''(19052006). We thank Supercomputer Center, ISSP,
University of Tokyo for providing us with the CPU time.

\newpage

\noindent
{\Large Supplemental Material}

\bigskip\bigskip
In this Supplemental Material, we explain some of the details about (I) the equilibration check of our Monte Carlo simulations, (II) the states of the chiral domain walls, (III) the properties of the $Z_2$ vortices, and (IV) the manner how the $Z_2$ vortices and the chiral domain walls interact with each other in our model.

\section{Equilibration check of Monte Carlo simulations}

In this section, we explain how we check equilibration in our Monte Carlo (MC) simulations. Since our simulations are performed at very low temperatures, an appropriate check of equilibration is crucially important. The standard check might be to monitor the stability of various observables as a function of MC time for successively longer observation times, which we do perform.  In addition to this standard test, we also perform the following procedure for an additional check of equilibration. 

\begin{figure}[h]
  \includegraphics[width=8cm]{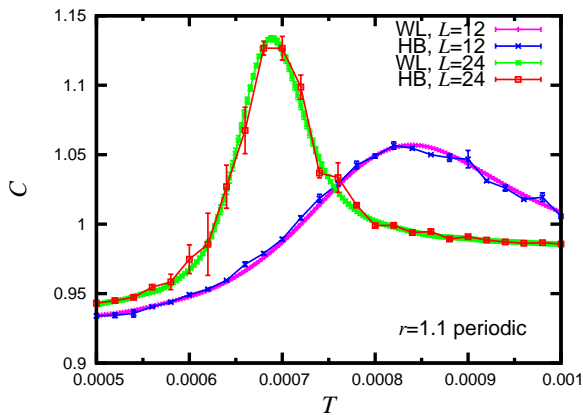}
 \caption{
 The temperature dependence of the specific heat calculated either by the heat-bath (HB) method combined with the over-relaxation method, or by the Wang-Landau (WL) method. The distortion parameter is $r=1.1$. 
}
\end{figure}

 As mentioned in the main text, we take our data based on the combination of the heat-bath (HB) method and the over-relaxation method. (The temperature-exchange technique often useful in, {\it e.g.\/}, spin-glass problems, is not very efficient here at such low temperatures.) To be sure that our data are fully thermalized, we also take the data of the same physical quantities based on the so-called Wang-Landau method \cite{WL} and check whether the data taken by the two different methods agree within the errors. Note that the Wang-Landau method directly computes the density of states, and the temperature comes into physical quantities in a way very different from the heat-bath (+over-relaxation) method. Thus, the agreement of the datasets taken by these two different methods should yield a stringent test of equilibration. In Fig.5, we show the temperature dependence of the specific heat for the case of $r=1.1$ calculated either by the heat-bath (+ over-relaxation) method or by the Wang-Landau method for the sizes $L=12$ and 24. As can be seen from the figure, the two kinds of datasets agree completely, demonstrating that our data are a fully thermalized even at the lowest temperature studied. For larger sizes, the Wang-Landau method is very much time-consuming, and we are unable to obtain the data for larger sizes. Yet, even for smaller sizes of $L=12$ and 24 shown here, the existence of the growing specific-heat peak is already clear, demonstrating the occurrence of an equilibrium phase transition at this very low temperature.

\section{Chiral domain walls} 

In this section, we show how the chiral domain walls are thermally generated  in our model both above and below $T_c$. In Fig.6, we give typical snapshots of the chiral domain-wall configurations  below $T_c$, $T=5\times 10^{-4}$ [left figure] and above $T_c$, $T=7\times 10^{-4}$ [upper figure] for a system of the distortion parameter $r=1.1$ where $T_c\simeq 6.5\times 10^{-4}$. Free boundary conditions (BC) are applied here, because in general incommensurate cases periodic BC applied to finite lattices force the system to introduce a pair of chiral domain walls which are extended in the perpendicular direction. By contrast, no such constraint exists in free BC which we adopt here.  Chiral domain walls, defined  at a lattice point $i$ connecting two bonds ($i-\hat e_\mu, i$) and ($i, i+\hat e_\mu$) in line in the manner as explained in the main text, are depicted in the figure as blue segments perpendicular the bond and passing the site $i$.

\begin{figure}[h]
  \includegraphics[width=8cm]{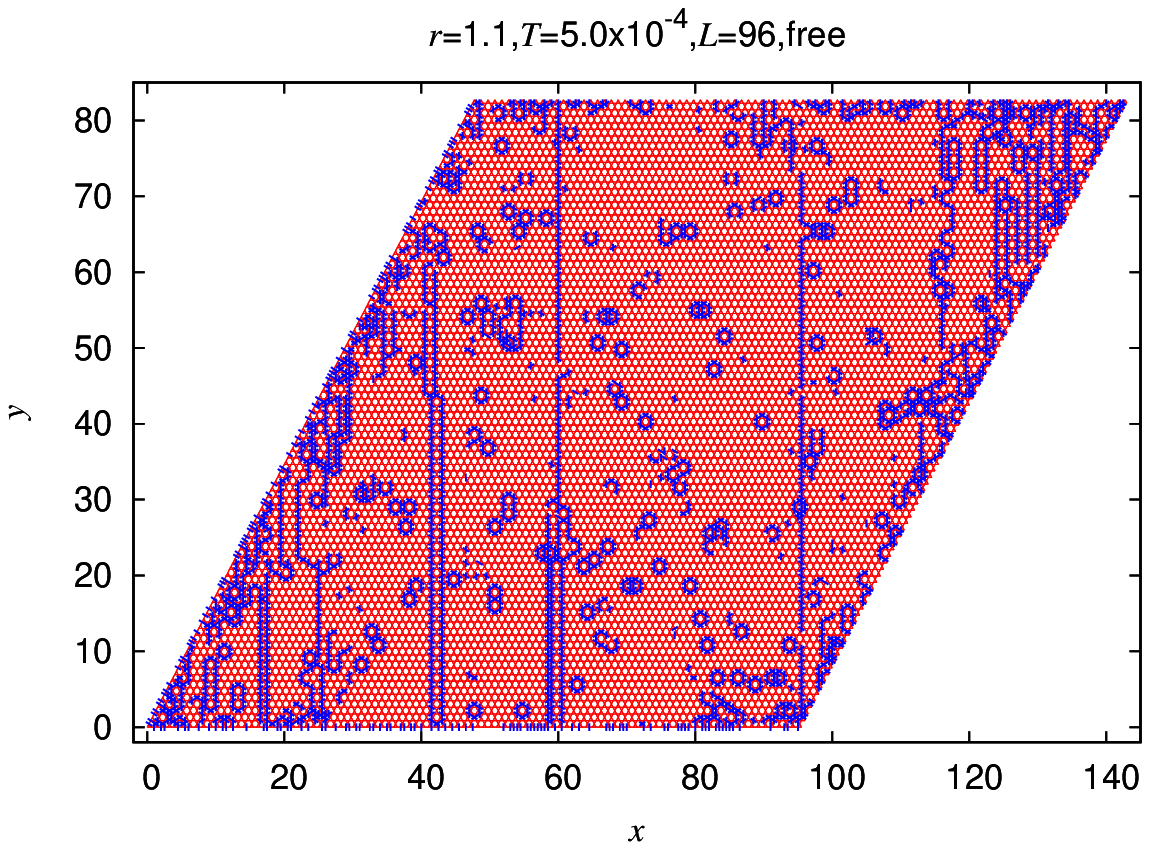}
  \includegraphics[width=8cm]{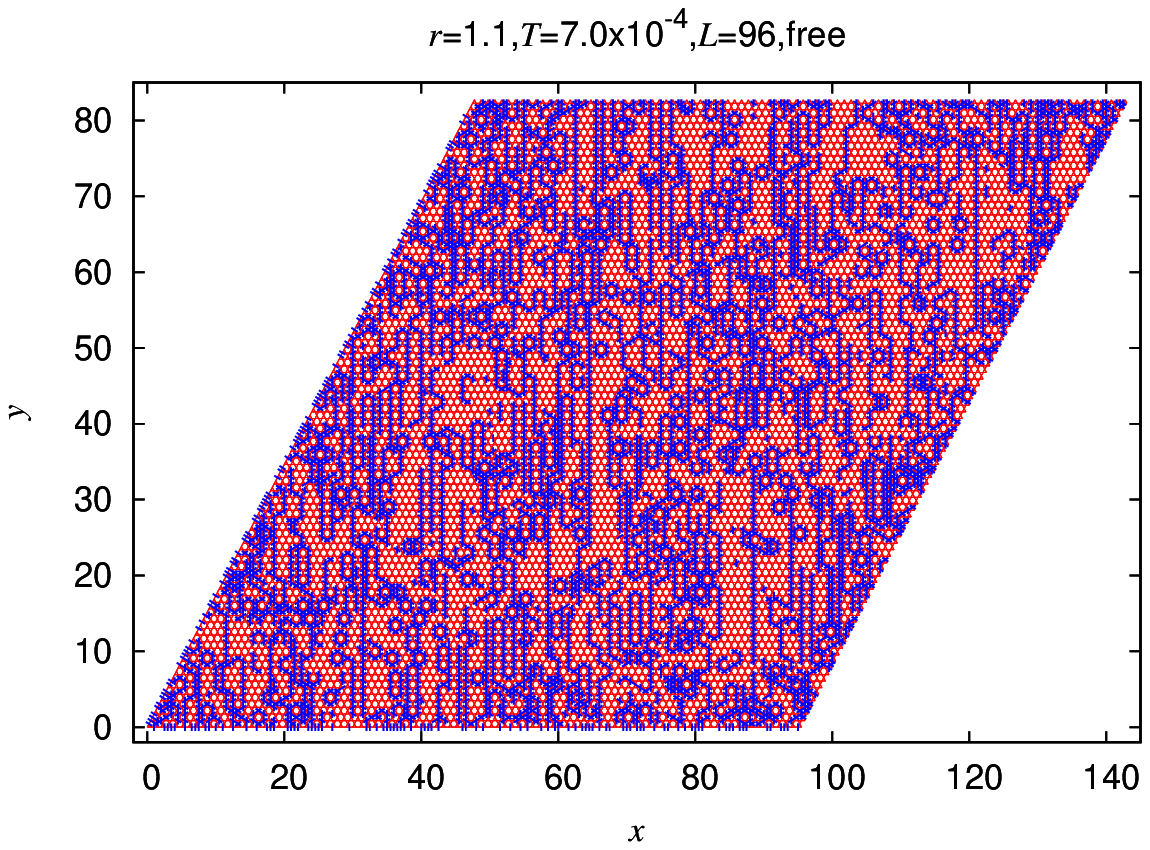}
 \caption{
Typical snapshots of the chiral domain-wall configurations both below $T_c$, $T=5\times 10^{-4}$ [upper figure], and above $T_c$, $T=7\times 10^{-4}$ [lower figure]. The distortion parameter is $r=1.1$. Free BC are applied for a system of linear size $L=96$. 
}
\end{figure}

 Chiral domain walls may be classified into two types: An extended wall spanning the entire lattice from one end to the other perpendicular to the $J_2$-bond direction and a closed-loop wall. Note that a closed-loop wall costs a small but nonzero energy, while a perfectly straight-line wall  perpendicular to the $J_2$-bond direction costs no energy (zero mode) owing to the large degeneracy of the ground state of the $J_1-J_2$ kagome Heisenberg model. As can be seen from the figure, both types of chiral domain walls are present below and above $T_c$, whereas their number density increases rapidly as the temperature is increased across $T_c$. As can be seen from the upper figure, an extended wall running along the perpendicular direction seems to persist in small number even below $T_c$ (note that free BC are applied here so that the  observed chiral domain wall is not the one forced by periodic BC). It may be no wonder that extended walls remain in small number even at low temperatures, since they are zero-energy modes so long as they run straight along the perpendicular direction.

\section{$Z_2$ vortices}

In this section, we explain some of the basic properties of the $Z_2$ vortex, together with the way how we identify the spatial position of the $Z_2$ vortex (more precisely, the position of its core) in our simulations. The $Z_2$ vortex is a topologically stable point defect inherent to two-dimensional frustrated Heisenberg systems with the noncollinear spin order. It possesses a parity-like $Z_2$ ($\pm 1$) topological quantum number depending on whether there is a vortex or no vortex. As such, a winding number, familiar as a good topological quantum number characterizing the standard vortex, is not a good topological quantum number here.

 The properties of the $Z_2$ vortices were discussed in some detail in Ref.[2] in the context of the classical Heisenberg antiferromagnet (AF) on the triangular-lattice with the nearest-neighbor coupling. The $Z_2$ vortex might be regarded as a vortex formed by the chirality vector. It was suggested in Refs.\cite{KM,KYO} that the triangular Heisenberg AF might exhibit a thermodynamic phase transition at a finite temperature driven by the binding-unbinding of the $Z_2$ vortices,  keeping the standard spin correlation length to be finite.  In the previous studies, as an order parameter characterizing the $Z_2$-vortex transition, either a Wilson-loop (a vorticity function)  \cite{KM} or a vorticity modulus \cite{KK,KYO} has been proposed. The vorticity modulus $v$ might be defined as the total free-energy cost against a vortex formation $\Delta F_{tot}$ divided by $\ln L$, $v=\Delta F_{tot}/\ln L$. In the triangular-lattice Heisenberg AF in its thermodynamic limit, $v$ should become zero in the high-temperature phase where there exist free vortices, while $v$ should take a nonzero value in the low-temperature phase where all vortices are paired with no free vortices. In case of our kagome model, the existence of the chiral domain-wall excitations would modify the nature of the vortex transition as we shall see in the next section.

 In our simulations on a (distorted) kagome Heisenberg AF, we monitor the spatial position of the $Z_2$ vortex following the procedures of Ref.\cite{KM}. The triangular lattice consists of three interpenetrating triangular sublattices, and one might take an elementary plaquette a minimum upward triangle consisting of the three sublattice sites.  At low temperatures, the local spin structure at each plaquette is approximately a $120$-degrees structure. In Ref.\cite{KM}, an $SO(3)$ local `frame' was assigned to each such elementary plaquette, and a $Z_2$ vortex is identified for every minimum triangular loop consisting of three such plaquettes. We modify such a procedure of the local vortex identification to our kagome system.

 First, an elementary plaquette is taken again as a minimum upward triangle consisting of three spins, which coincides with a crystallographic unit cell of the kagome lattice.  The lattice formed by these elementary plaquettes also forms the triangular super-lattice consisting of the three triangular sublattices A, B and C. The spin structure here is an incommensurate spiral state close to the $\sqrt 3\times \sqrt 3$ state, as demonstrated in the main text.  In the perfect $\sqrt 3\times \sqrt 3$ state, the 120-degrees structure on an elementary plaquette would exhibit a uniform arrangement on each sublattice A, B and C. In view of this, we assign our local `frame' to an elementary plaquette on one of these three sublattices, say, on A, and identify an elementary $Z_2$ vortex for every minimum triangular loop consisting of three such plaquettes of the sublattice A. While the true spin structure generally deviates from the $\sqrt 3\times \sqrt 3$ structure, this deviation hardly affects the definition of the $Z_2$ vorticity since only the topological character matters in the definition. In computing the spatial distribution of the $Z_2$ vortices in the following section, we use this definition of the local $Z_2$ vorticity.

\begin{figure}[h]
  \includegraphics[width=9cm]{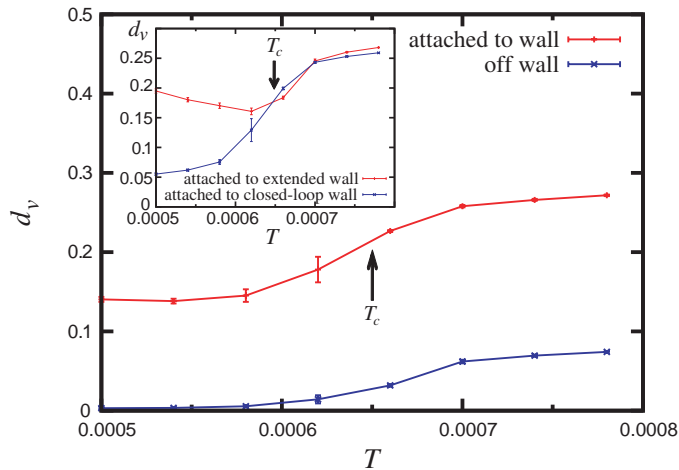}
 \caption{
The temperature dependence of the number density (the appearance probability) of $Z_2$ vortices for the two kinds of sites, the site attached to the chiral domain wall and the site away from the chiral domain wall. The distortion parameter is $r=1.1$, periodic BC being applied. The inset represents the temperature dependence of the number density of the $Z_2$ vortices for the two kinds of chiral domain-wall sites, the site attached to an extended, system-spanning wall, and the site attached to a closed-loop wall. The arrow indicates the transition point.
}
\end{figure}

\section{$Z_2$ vortices and chiral domain walls}

In this section, we explain the manner how the $Z_2$ vortices and the chiral domain walls interact with each other both above and below $T_c$.

In Fig.7, we show the temperature dependence of the number density (or the appearance probability) of $Z_2$ vortices, $d_v$, for the two kinds of sites, the site attached to the chiral domain wall and the site away from the chiral domain wall. The distortion parameter is set $r=1.1$. As can immediately be seen from the figure, the sites attached to the chiral domain wall allow higher density of $Z_2$ vortices. Namely, the $Z_2$ vortices are generated primarily at or near the chiral domain walls. It means that there exists an effective {\it attractive\/} interaction between the $Z_2$ vortices and the chiral domain walls. The inset represents the temperature dependence of the number density of $Z_2$ vortices for the two kinds of chiral domain-wall sites, the site attached to an {\it extended\/}, system-spanning domain wall and the site attached to a {\it closed-loop\/} wall. One can see from the inset that  at lower temperatures an extended chiral domain wall accommodates a $Z_2$ vortex preferably.

\begin{figure}[t]
  \includegraphics[width=7.4cm]{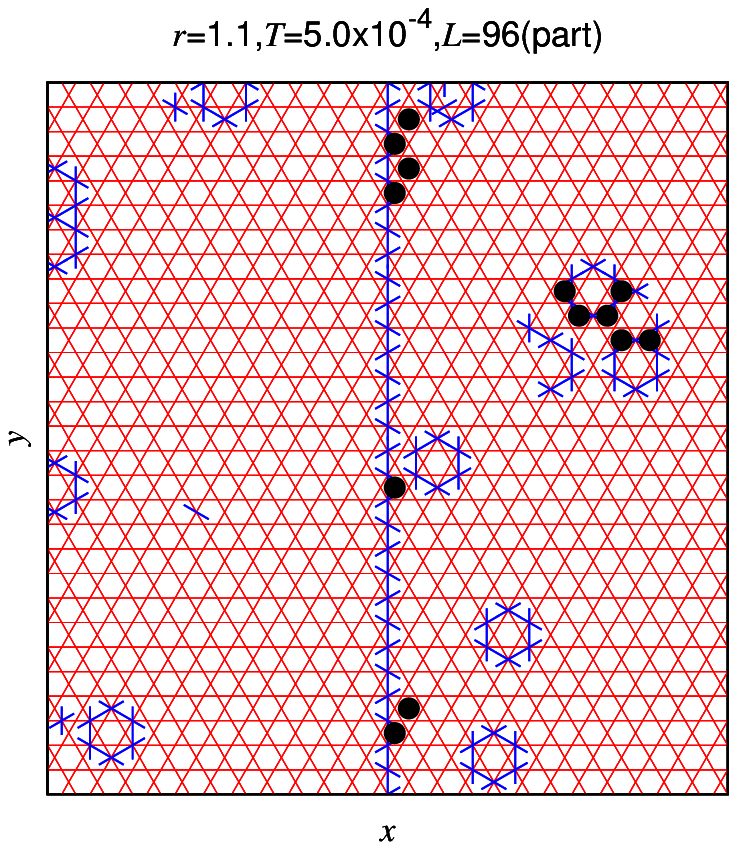}
  \includegraphics[width=7.4cm]{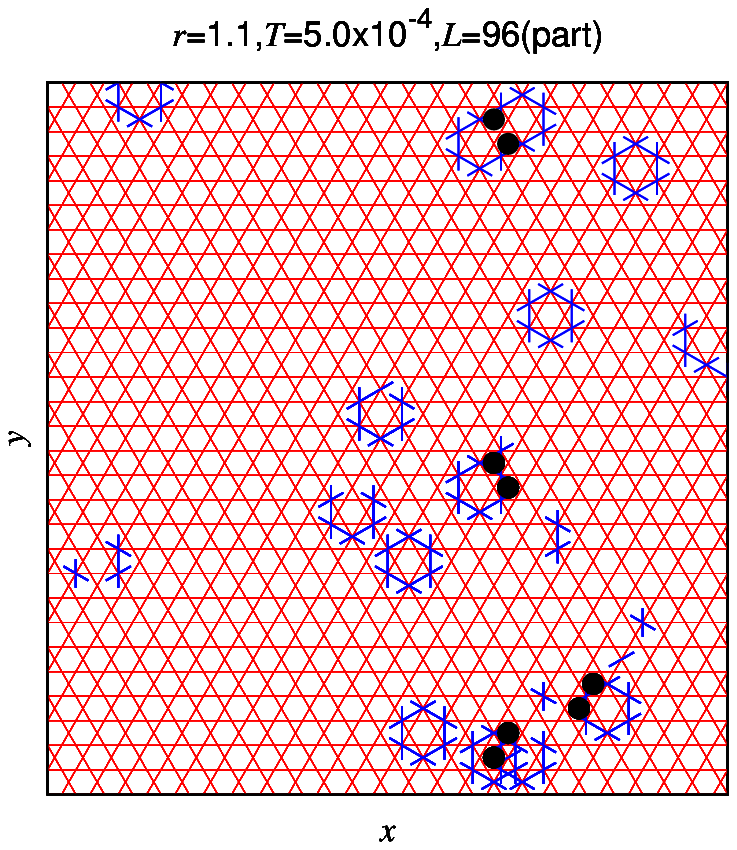}
  \includegraphics[width=7.4cm]{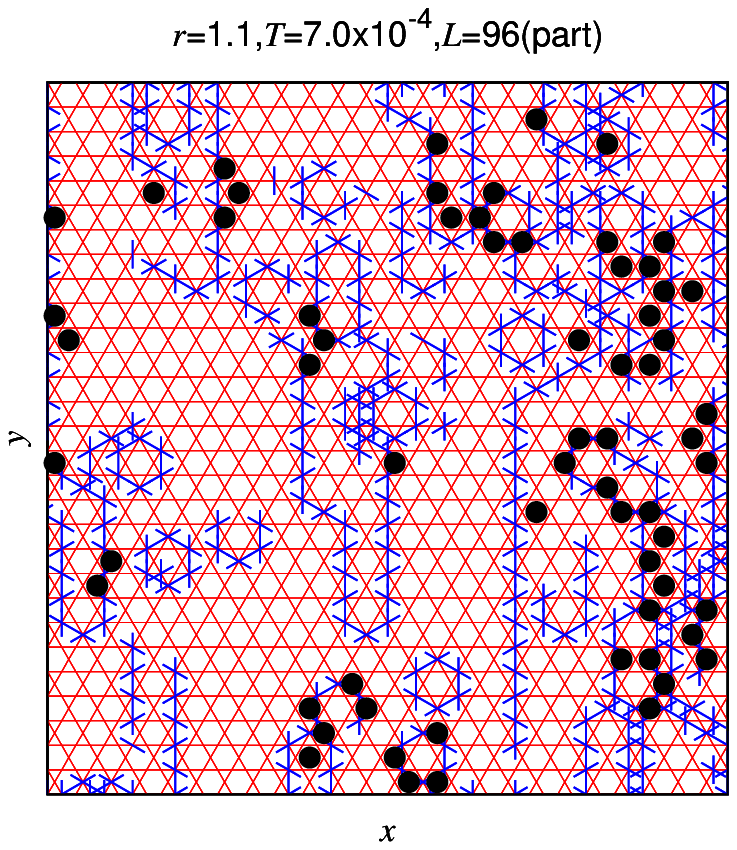}
 \caption{
Snapshots of the $Z_2$-vortex configurations both below $T_c$, $T=5\times 10^{-4}$ [top and middle figures], and above $T_c$, $T=7\times 10^{-4}$ [bottom figure], depicted by black dots. Chiral domain wall are also shown by blue lines. The distortion parameter is $r=1.1$, periodic BC being applied. All figures are parts of the $L=96$ lattice. The top figure is a part of the lattice containing an extended chiral domain wall, whereas the middle figure is another part of the same lattice not containing an extended chiral domain wall, {\it i.e.\/}, containing closed loops only. Below $T_c$, while an extended wall allows isolated, apparently free $Z_2$ vortices, $Z_2$ vortices appear always in bound pairs at closed-loop wall sites.  
}
\end{figure}

 We note that the existence of an effective attractive interaction acting between the chiral domain walls and the $Z_2$ vortices might provide a possible explanation of the observed first-order transition. As the temperatures is increased, both the chiral domain walls and the $Z_2$ vortices tend to be thermally excited. If there is a strong attractive interaction between these two topological excitations as we have demonstrated above, one would expect a positive feed-back effect operative in their thermal generation: {\it i.e.\/}, the generation of one induces the generation of the other. Thus, the observed first-order transition might be a consequence of such {\it cooperative\/} thermal generation of the chiral domain walls and the $Z_2$ vortices.

 In Fig.8, we show snapshots of the $Z_2$-vortex configurations both below $T_c$, $T=5\times 10^{-4}$ [top and middle figures], and above $T_c$, $T=7\times 10^{-4}$ [bottom figure], together with the corresponding chiral domain-wall configurations. The distortion parameter is set $r=1.1$. All figures are parts of the $L=96$ lattice. The top and the middle figures represent different parts of the same lattice measured at exactly the same MC time. The top figure is a part of the lattice containing an extended chiral domain wall, whereas the middle figure is a part not containing an extended chiral domain wall, {\it i.e.\/}, a part containing closed-loop walls only. As can be seen from the top figure, an extended wall allows isolated, apparently free $Z_2$ vortices even below  $T_c$ in addition to $Z_2$-vortex pairs, whereas closed-loop walls bear only $Z_2$-vortex pairs below $T_c$. Above $T_c$,  as can seen from the bottom figure, free vortices are generated not only at extended-wall sites but also at closed-loop wall sites, whereas the number of $Z_2$ vortices becomes large and the distinction between free vortices and vortex-pairs become somewhat obscure here.

\end{document}